\documentclass[aps,pre,epsf,superscriptaddress,amsmath,amssymb,amsfonts,twocolumn,showpacs]{revtex4-1}
\usepackage{amsmath,amssymb,amsfonts,bm,comment}
\usepackage{graphics,float,tikz}
\usetikzlibrary{arrows.meta,arrows}
\usepackage[caption=false]{subfig}
\usepackage[colorlinks=true,
	linkcolor=blue,
	filecolor=magenta,      
	urlcolor=blue,
	citecolor=blue]{hyperref}
\usepackage{physics,xparse,mathtools,isotope}
\usepackage[table-parse-only]{siunitx}
\usepackage{multirow,array,relsize,lipsum,longtable,tabularx}
\usepackage{dsfont}

\newcommand{\gniewko}[1]{\color{red}{#1}\color{black}}
\newcommand{\madhura}[1]{\color{blue}{#1}\color{black}}

\newtheorem{conjecture}{Conjecture}

\begin{document}
	\title{Detecting Entanglement Between Modes of Light}

	\author{Madhura Ghosh Dastidar}
	\affiliation{Department of Physics, Indian Institute of Technology Madras, Chennai 600036, Tamil Nadu, India}
	
	\author{Gniewomir Sarbicki}
	\affiliation{Institute of Physics, Faculty of Physics, Astronomy and Informatics, Nicolaus Copernicus University, Grudzi\c adzka 5/7, 87-100 Toru\'n, Poland}
	\date{\today}
	
	\begin{abstract}
		
    We consider a subgroup of unitary transformations on a mode of light induced by a Mach-Zehnder Interferometer and an algebra of observables describing a photon-number detector proceeded by an interferometer. We explore the uncertainty principles between such observables and their usefulness in performing a Bell-like experiment to show a violation of the CHSH inequality, under physical assumption that the detector distinguishes only zero from non-zero number of photons. We show which local settings of the interferometers lead to a maximal violation of the CHSH inequality.

	\end{abstract}
	\maketitle
	
	\section{Introduction}\label{Introduction}
	
	Multiphoton entangled states~\cite{caspani2017integrated,dell2006multiphoton} have applications in the fields of quantum communications~\cite{qin2014multiphoton}, computation~\cite{langford2011efficient, walther2005experimental} and metrology~\cite{paulisch2019quantum,afek2010high}. Aside from polarisation, optical modes of photons are also another property that can be entangled, as seen for Dicke superradiant photons~\cite{paulisch2019quantum}. 
	Such quantum states consist of many photons that may be mode-entangled.
	
	Hilbert space of $n$ photons, which can be in two modes (of polarisation or wave vector), is a symmetric subspace (due to bosonic nature of photons) of $(\mathds{C}^2)^{\otimes n}$. If the number of photons in experiment is not known, then we deal with direct sum of such spaces with different $n$s (Fock space). On the other hand, we can consider a quantum state of light consisting two modes, each being occupied by an arbitrary number of photons. Such state lives on tensor product of the Hilbert spaces of modes: $\mathcal{H}_1 \otimes \mathcal{H}_2$, and may be entangled in general. Further, if we consider two optical modes, the more natural approach is not to have any restrictions on the number of photons. If the number of photons is fixed, then a state of light is supported in an eigenspace of global photon number: $\hat N \otimes \mathds{I} + \mathds{I} \otimes \hat N$, being isomorphic to the symmetric sector of $(\mathbb{C}^2)^{\otimes n}$. The whole Hilbert space $\mathcal{H}_1 \otimes \mathcal{H}_2$ is isomorphic to the whole Fock space. In this paper we will consider entanglement of a quantum state of two modes of light each being occupied by an arbitrary number of photons. 
	
	In general, entanglement can be detected by estimating the density matrix of the quantum state of the system~\cite{white1999nonmaximally,schwemmer2014experimental} and mathematically testing for its non-separability using various separability criteria~\cite{horodecki2009quantum}. However, reconstruction of the entire density matrix via quantum state tomography~\cite{paris2004quantum} with many photons in each mode is challenging due to the large number of entries of the density matrix, each requiring many measurements to obtain a desired accuracy.
	Another approach is to measure an expected value of appropriately chosen entanglement witness~\cite{chruscinski2014entanglement} and estimate only one parameter instead of all entries of density matrix. 
	
	Bell inequality~\cite{bell1964einstein,maccone2013simple} is an algebraic expression built from local observables satisfying certain assumptions. Expected value of such expression satisfies a certain bound for all separable states. Fixing these observables one obtains an entanglement witness~\cite{hyllus2005relations}.
	
	The most famous Bell inequality is the CHSH inequality~\cite{clauser1969proposed}: $\mathds{E} ( A_1 \otimes B_1 + A_1 \otimes B_2 + A_2 \otimes B_1 - A_2 \otimes B_2 ) \le 2$. With appropriate choice of local observables, the CHSH inequality can be violated for certain entangled states with its LHS reaching the value of $2\sqrt{2}$ known as Tsirelson's bound~\cite{cirel1980quantum}.
	
	

    In section \ref{U section}, we consider an action on one mode state of light of a Mach-Zehnder Interferometer (MZI) fed with strong coherent state of light
    on its second input port. We show, than in the limit of strong coherent field the MZI setup acts as a unitary operation on an input state and hence a projective measure of its output corresponds to a projective measurement on its input. However, for finite coherent fields the action of MZI setup is rather that of a quantum channel and the resulting measurement on its input will be a POVM. In section \ref{POVMs}, we discuss how in this limit the quantum channel becomes a unitary transformation and the POVM becomes a projective measurement. We will proceed in the limiting scenario when the MZI realises a unitary transformation.
    
    Next, in section \ref{MU section}, we discuss the unitary operators related to the action of MZI and the algebra of observables representing photon number measurements proceeded by an interferometer.
    In particular, we discuss uncertainty relations between these observables.

    Finally, in section \ref{CHSH section} we discuss, how one can perform a Bell-like experiment measuring the violation of CHSH inequality in such a scenario. We show that, with appropriate setups of interferometers, we are able to obtain the maximum possible violation of CHSH inequality.
	\section{Unitary Transformations}\label{U section}
	
	For photons, optical components such as beam splitters and phase shifters can be used to generate unitary transformations in the cumulative Fock state.
	
	\subsection{Beam Splitter Implementation}\label{BS subsection}
	
	The effect of a beam splitter on a photonic state can be envisioned as a unitary operation on the incoming photon states. A typical "quantum" beam splitter schematic is shown in Fig.~\ref{BeamSplitter}. The photon annihilation operators at the output ports [$\hat{a}_{2}$, $\hat{a}_{3}$] corresponding to the respective input ports [$\hat{a}_{0}$, $\hat{a}_{1}$] are transformed as~\cite{windhager2011quantum}:
	
	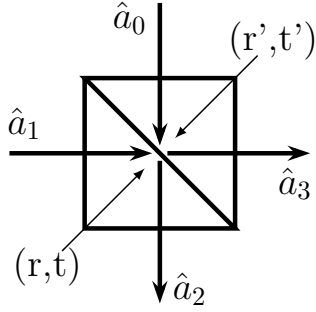
\begin{figure}[ht]
		\centering
		\begin{tikzpicture}[>=latex,line width = .6mm, font=\relsize{3}]
		    \draw (-1,1) -- (-1,-1) -- (1,-1) -- (-1,1) -- (1,1) -- (1,-1);
		    \draw[-{Stealth[length=3mm, width=2mm]}] (0,2) -- (0,.1);
		    \draw[-{Stealth[length=3mm, width=2mm]}] (0,-.1) -- (0,-2);
		    \draw[-{Stealth[length=3mm, width=2mm]}] (-2,0) -- (-.1,0);
		    \draw[-{Stealth[length=3mm, width=2mm]}] (.1,0) -- (2,0);
		    \draw[->,line width=.2mm] (1.3,1.3) -- (.2,.2);
		    \draw[->,line width=.2mm] (-1.3,-1.3) -- (-.2,-.2);
		    \node at (1.5,1.5) {(r',t')};
		    \node at (-1.5,-1.5) {(r,t)};
		    \node at (-.4,1.8) {$\hat a_0$};
		    \node at (1.8,-.4) {$\hat a_3$};
		    \node at (.4,-1.8) {$\hat a_2$};
		    \node at (-1.8,.4) {$\hat a_1$};
		\end{tikzpicture}
		\caption{\textbf{Quantum Beam Splitter.} Schematic diagram of a quantum beam splitter with two input ports ($\hat{a}_0$, $\hat{a}_1$) and two output ports ($\hat{a}_2$, $\hat{a}_3$), with corresponding reflectivities and transmittivities, ($r$, $t$) and ($r'$, $t'$) at the input and output ports, respectively.}
		\label{BeamSplitter}
	\end{figure}
	
	\begin{equation}
	    \label{BS: Matrix Representation}
	    \begin{pmatrix}
        \hat{a}_{2}\\
        \hat{a}_{3}
        \end{pmatrix} =
	    \begin{pmatrix}
        t' & r\\
        r' & t
        \end{pmatrix}\begin{pmatrix}
        \hat{a}_{0}\\
        \hat{a}_{1}
        \end{pmatrix} 
 	\end{equation}
	where ($r$, $t$)[($r'$, $t'$)] are the reflectance and transmittance of the beam splitter at the input[output] ports, respectively. Due to energy conservation, these numbers are complex in general and form a unitary matrix, i.e:
	\begin{eqnarray}
	    |t|^2 + |r|^2 = 1 \\
	    |t'|^2 + |r'|^2 = 1 \\
	    t'r^* + r't^* = 0 \label{Stokes: eqn 4}
	\end{eqnarray}
	It implies in particular, that $|t| = |t'|$ and $|r| = |r'|$. The above equations are often referred to as Stokes' laws.
	In general, for a single photon input, the beam splitter performs a rotation on the Poincare sphere~\cite{leonhardt2003quantum}.

	Consider a general many-photon Fock state:
	\begin{equation}
	    \label{general fock state}
	    \ket{\psi} = \sum_{n=0}^{\infty} c_n \ket{n} = \sum_{n=0}^{\infty} c_n\frac{1}{\sqrt{n!}}(\hat{a}_{0}^\dagger)^n \ket{0}
	\end{equation}
	and a coherent state of light $\hat{D}_1(\alpha) \ket{0}_1$, where
	\begin{equation}
	    \hat{D}_1(\alpha) = \mathrm{exp} (\alpha\hat{a}^\dagger_{1}-\alpha^*\hat{a}_{1})
	\end{equation}
	is the displacement operator,
	to be incident on the first and second port of the beam splitter, respectively. The total input state of the BS is
	
	\begin{equation}
	    \label{BS: Input state}
	     \ket{\psi}_0\otimes\ket{\alpha}_1 = \sum_{n=0}^{\infty} c_n\frac{1}{\sqrt{n!}}(\hat{a}_{0}^\dagger)^n \ket{0}_0 \otimes \hat{D}_1(\alpha) \ket{0}_1 
	\end{equation}

	Assuming the beam splitter operator to be $\hat{U}_1$ from (\ref{BS: Matrix Representation}), we get the photon annihilation operators ($\hat{a}_0$ and $\hat{a}_1$) in terms of that at the output ports ($\hat{a}_2$ and $\hat{a}_3$) as,
	
	\begin{equation}
	    \label{output photon operators}
	    \begin{gathered}
	        \hat{a}_0 = t'^*\hat{a}_2 + r'^*\hat{a}_3
	        \quad \textrm{and} \quad
	        \hat{a}_1 = r^*\hat{a}_2 + t^*\hat{a}_3
	    \end{gathered}
	\end{equation}
	
	where we have used the Stokes' laws: $r^*t'+r't^*=0$ and $\abs{r}^2+\abs{t}^2=1$ along with [Eq.~\ref{BS: Matrix Representation}]. 
	
	Applying the beam splitter (BS) transformation (\ref{BS: Matrix Representation}) to operators in the input state formula (\ref{BS: Input state}) we obtain
	
    \begin{widetext}
	\begin{equation}\label{BS: Output state}
	\begin{gathered}[b]
	    \ket{\psi}_0 \otimes \ket{\alpha}_1 \xrightarrow[]{BS} \ket{\Psi}_{out} = 
	    \mathrm{exp}(\alpha(r\hat{a}_{2}^\dagger + t\hat{a}_{3}^\dagger)-\alpha^*(r^*\hat{a}_{2} + t^*\hat{a}_{3}))\sum_{n=0}^{\infty} \frac{c_n}{\sqrt{n!}}(t'\hat{a}_{2}^\dagger + r'\hat{a}_{3}^\dagger)^n
	    \ket{0}_2\otimes\ket{0}_3 \\
	    = \mathrm{exp}(r\alpha\hat{a}_{2}^\dagger - r^*\alpha^*\hat{a}_{2}) \mathrm{exp}(t\alpha\hat{a}_{3}^\dagger-t^*\alpha^*\hat{a}_{3})\sum_{n=0}^{\infty} c_n\frac{1}{\sqrt{n!}}(t'\hat{a}_{2}^\dagger + r'\hat{a}_{3}^\dagger)^n
	    \ket{0}_2\otimes\ket{0}_3 \\
	    = \hat{D}_2(r\alpha) \hat{D}_3(t\alpha)\sum_{n=0}^{\infty}c_n\frac{1}{\sqrt{n!}}(A^\dagger)^n \ket{0}_2\otimes\ket{0}_3
	\end{gathered}
	\end{equation}
\end{widetext}



where $A^\dagger = t'\hat{a}_{2}^\dagger + r'\hat{a}_{3}^\dagger$. In the limit of a highly reflective beam splitter and a highly intense coherent state:
\begin{equation}
    \label{BS: limit}
    r \xrightarrow[]{} 1, \quad t \alpha = \mathrm{const.}
\end{equation}
the output state formula (\ref{BS: Output state}) reduces to:

\begin{equation}
\label{BS: Output state 1}
    \ket{\Psi}_{out} = \ket{r\alpha}_2 \otimes \hat{D}_3(t\alpha) \ket{\psi}_3
\end{equation} 

Thus, we achieve the incoming coherent state with reduced intensity ($\ket{r\alpha}_2$) and the incoming photonic state displaced by $t\alpha$ at the output ports 2 and 3 respectively.
Using these results whereby the beam splitter displaces any quantum state, one can physically implement unitary transformations over the photonic wavepacket. However in this case, the parameters of displacement, i.e., $t$ and $\alpha$ depend only on the transmittivity of the beam splitter and the input coherent field intensity, respectively. Moreover, a highly reflective beam splitter with $r \xrightarrow[]{} 1$ is practically difficult to construct. To eliminate such problems with the implementation of the scheme and to exercise further degree of tunability on the displacement operator, we describe the case of using a MZI setup with the same input state (see [Eq.~\ref{BS: Input state}]).

\subsection{Mach-Zehnder Interferometric Implementation} \label{MZI section}

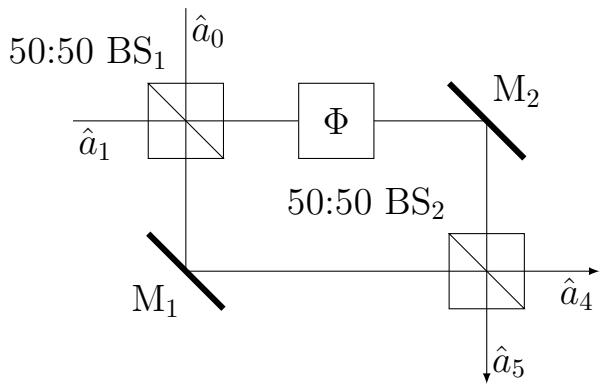
\begin{figure}[ht]
		\centering
		\begin{tikzpicture}[>=latex,font=\relsize{3}]
		    \draw (0,1) -- (0,0) -- (1,0) -- (0,1) -- (1,1) -- (1,0);
		   \draw[shift={(4,-2)}] (0,1) -- (0,0) -- (1,0) -- (0,1) -- (1,1) -- (1,0);
		    \draw[shift={(0,-2)}, line width=.8mm] (0,1) -- (1,0);
		    \draw[shift={(4,0)}, line width=.8mm] (0,1) -- (1,0);
		    \draw[->] (-1,.5) -- (4.5,.5) -- (4.5,-3);
		    \draw[->] (.5,2) -- (.5,-1.5) -- (6,-1.5);
		    \draw[fill=white,shift={(2,0)}] (0,1) rectangle (1,0);
		    \node at (2.5,.5) {$\Phi$};
		    \node at (-.8,1.4) {50:50 BS\textsubscript{1}};
		    \node at (2.9,-.6) {50:50 BS\textsubscript{2}};
		    \node at (.1,-1.9) {M\textsubscript{1}};
		    \node at (4.9,.9) {M\textsubscript{2}};
		    \node at (-.7,.2) {$\hat a_1$};
		    \node at (.8,1.7) {$\hat a_0$};
		    \node at (4.8,-2.7) {$\hat a_5$};
		    \node at (5.7,-1.8) {$\hat a_4$};
		\end{tikzpicture}
		\caption{\textbf{Mach-Zehnder Interferometer.} MZI setup with a phase shift ($\phi$) in one of the arms of the interferometer. Two 50:50 beam splitters (BS$_1$ and BS$_2$, with BS$_2$ $180^o$ rotated w.r.t BS$_1$), i.e., having equal magnitudes of reflectivity and transmittivity, are used along with two mirrors (M$_1$ and M$_2$) for such an interferometer.}
		\label{MZI setup}
\end{figure}

A MZI can be approximated as a four-port device~\cite{yurke19862} as shown in Fig.~\ref{MZI setup}. The composite optical elements of the MZI setup each correspond to a unitary operation over the field states. Defining the matrix associated to the effect of the phase shifter over the input state as,

\begin{equation}
    \label{MZI: Phase Shifter}
    P_\phi =
    \begin{pmatrix}
    1 & 0\\
    0 & e^{i\phi}
    \end{pmatrix} 
\end{equation}

Using the definition of the beam splitter operator from [Eq.~\ref{BS: Matrix Representation}], we find the transformation of the annihilation operators to be,

\begin{equation*}
    \label{MZI: Matrix Representation}
    \begin{gathered}[b]
    \begin{pmatrix}
    \hat{a}_{4}\\
    \hat{a}_{5}
    \end{pmatrix} =
    \begin{pmatrix}
    t_2' & r_2\\
    r_2' & t_2
    \end{pmatrix} \begin{pmatrix}
    1 & 0\\
    0 & e^{i\phi}
    \end{pmatrix}
    \begin{pmatrix}
    t_1' & r_1\\
    r_1' & t_1
    \end{pmatrix}\begin{pmatrix}
    \hat{a}_{0}\\
    \hat{a}_{1}
    \end{pmatrix}\\
    \end{gathered}
\end{equation*}

Now, assuming that two identical beam splitters are arranged in the MZI setting such that the first beam splitter is aligned in the reverse direction relative to the second as shown in Fig.~\ref{MZI setup}, we have,

\begin{equation}\label{MZI: Matrix Opp BS}
    \begin{gathered}[b]
    \begin{pmatrix}
    \hat{a}_{4}\\
    \hat{a}_{5}
    \end{pmatrix} =
    \begin{pmatrix}
    t' & r\\
    r' & t
    \end{pmatrix} \begin{pmatrix}
    1 & 0\\
    0 & e^{i\phi}
    \end{pmatrix}
    \begin{pmatrix}
    t'^{*} & r^{*}\\
    r'^{*} & t^{*}
    \end{pmatrix}\begin{pmatrix}
    \hat{a}_{0}\\
    \hat{a}_{1}
    \end{pmatrix}\nonumber\\
    = \begin{pmatrix}
    \abs{t'}^2+\abs{r}^2e^{i\phi} & r'^*t'(1-e^{i\phi})\\
    r't'^*(1-e^{i\phi}) & \abs{r'}^2+\abs{t}^2e^{i\phi}
    \end{pmatrix}\begin{pmatrix}
    \hat{a}_{0}\\
    \hat{a}_{1}
    \end{pmatrix},
    \end{gathered}
\end{equation}
using [Eq.~\ref{Stokes: eqn 4}]. Now, assume both to be 50:50 beam splitters, i.e., $\abs{t}=\abs{t'}=\abs{r}=\abs{r'}=\frac{1}{\sqrt{2}}$. Also since all coefficients of reflection and transmission are complex numbers we can write, $r'=\abs{r'}e^{i\gamma_1}$ and $t'=\abs{t'}e^{i\gamma_2}$. Therefore, the above equation reduces to:

\begin{equation}
    \label{MZI: 50:50 BS rep}
    \begin{pmatrix}
    \hat{a}_{4}\\
    \hat{a}_{5}
    \end{pmatrix} = 
    \frac 12
    \begin{pmatrix}
    1+e^{i\phi} & 
    e^{i\gamma}(1-e^{i\phi})\\
    e^{-i\gamma}(1-e^{i\phi}) & 
    1+e^{i\phi}
    \end{pmatrix}\begin{pmatrix}
    \hat{a}_{0}\\
    \hat{a}_{1}
    \end{pmatrix}\\
\end{equation}

where $\gamma = \gamma_2 - \gamma_1$. 
Alternating roles of $\hat a_4$ and $\hat a_5$ one gets:

\begin{equation}
    \label{MZI: scattering matrix rep}
    \begin{pmatrix}
    \hat{a}_{5}\\
    \hat{a}_{4}
    \end{pmatrix} = \begin{pmatrix}
    T' & R\\
    R' & T
    \end{pmatrix}
    \begin{pmatrix}
    \hat{a}_0\\
    \hat{a}_1
    \end{pmatrix},
    \\
\end{equation}
where:
\begin{equation}\label{MZI: Reflectivity and Transmittivity 2}
\begin{gathered}[b]
    R = R' = \frac{1+e^{i\phi}}{2}, \\
    T = \frac{e^{i\gamma}(1-e^{i\phi})}{2}, \\
    T' = \frac{e^{-i\gamma}(1-e^{i\phi})}{2}.
\end{gathered}
\end{equation}

Thus the MZI scattering matrix is equivalent to that of a beam spitter with tunable parameters, namely, effective reflectivities ($R$ and $R'$)s and transmitivities ($T$ and $T'$).

In the limit $\phi\xrightarrow[]{}0$, we can use the Taylor expansion of $e^{i\phi}$ up to second term such that $1-e^{i\phi} \simeq -i\phi$. So [Eqs.~\ref{MZI: Reflectivity and Transmittivity 2}] modify to,

\begin{eqnarray}
\label{MZI: R and T final}
    \lim_{\phi\to0}R = \lim_{\phi\to0}\frac{2+i\phi}{2} \simeq 1 \\
    \lim_{\phi\to0}T = \lim_{\phi\to0}-\frac{i\phi}{2}e^{i\gamma} \simeq 0
\end{eqnarray}

Drawing an analogy to Sec.~\ref{BS subsection}, we would require $T\alpha$ to remain constant (see [Eq.~\ref{BS: limit}]). For this: $|\alpha| \sim 1/
\phi$. Proportionality constant and phase of $\alpha$ will establish a proper displacement in [Eq.~\ref{BS: Output state 1}]. We are able to displace the input quantum state by $T\alpha$ using a MZI setup with two identical 50:50 beam splitters and small phase difference between arms, fed with strong laser field in a coherent state.


In case of $\alpha$ being finite, the state of the outputs is weakly entangled and tends to a separable state when $\alpha \to \infty$.

\section{POVM\lowercase{s}} \label{POVMs}


Let us assume from now, that the bottom arm of the MZI setup is ended by a photon number detector i.e. we measure intensity of field represented by the photon number operator $\hat N = \hat a^\dagger \hat a$. The MZI setup applies a global unitary transformation on the product state of the composite system (coherent state + multiphoton state). Let the projectors on the coherent and multiphoton states be $\sigma = \ketbra{\alpha}{\alpha}$ and $\rho = \ketbra{\psi}{\psi}$, respectively.

Thus, 
the state of the outputs of the MZI setup fed with an input state $\sigma\otimes\rho$ is,

\begin{equation}
    \mathrm{\varepsilon} = \mathrm{U}(\sigma\otimes\rho)\mathrm{U}^\dagger
\end{equation}
where $\mathrm{U}$ is the unitary operation performed by the MZI setup. Note that $\mathrm{U}$ has infinite dimensionality in the Schr\"odinger picture. The blocks of matrix $\mathrm{\varepsilon}$ are:

\begin{equation}\label{Emn}
    \mathrm{\varepsilon_{mn}} =
    \mathrm{Tr_1}(\varepsilon \ketbra{m}{n} \otimes \mathds I)
    =
    \sum_{ij}\mathrm{U_{mi}}(\sigma_{ij}\rho)(\mathrm{U_{nj}})^\dagger
\end{equation}
where $\mathrm{U_{mi}}$ is the $mi$-th block of $U$.

In the standard Fock basis, $\sigma_{ij}=\braket{i}{\alpha}\braket{\alpha}{j}\equiv\alpha_i\alpha^*_j$. Substituting the same in [Eq.~\ref{Emn}] and evaluating the trace of this quantum operation w.r.t the first subsystem ($\sigma$) we obtain:

\begin{align}
    \mathrm{\varepsilon} &= \sum_{k} \sum_{ij}(\alpha_i\mathrm{U_{ki}})\rho(\alpha_j\mathrm{U_{kj}})^\dagger
    = \sum_{k}E\rho E_k^\dagger
\end{align}
where $E_k = \sum_{i}\alpha_i\mathrm{U}_{ki}$ act as Kraus operators. It can be easily checked that these operators satisfy the completeness relation $\sum_k E_k E_k^{\dagger} = \mathds {I}$ (trace preservation).

The probability of observing $i$ photons at the detector is,

\begin{align}
    p(i) &= Tr(\ketbra{i}{i}\sum_{k}(\sum_j \alpha_j\mathrm{U_{kj}})\rho(\sum_j \alpha_j\mathrm{U_{kj}})^\dagger)) \nonumber \\
    &\equiv Tr(M_i\rho)
\end{align}

where $M_i = \sum_k (\alpha_j \mathrm{U_{kj}})^\dagger \ketbra{i} (\alpha_j \mathrm{U_{kj}})$
is the effect corresponding to measuring $i$ photons at the output. Using [Eq.~\ref{BS: Output state}], we derive this effect for the MZI setting to be:

\begin{widetext}
\begin{align}\label{Mi}
    M_i = e^{-\abs{T\alpha}^2} \frac{|T\alpha|^{2i}}{i!}
    \sum_{n,n'=0}^{\infty}
    &
    (-R'^*T\alpha)^{n'} (-R'^*T\alpha)^{*n}
    \sum_{k=0}^{\min\{n,n'\}} \frac{|R'\alpha|^{-2k}}{k!} \frac{\sqrt{n'!n!}}{(n'-k)!(n-k)!}
    \nonumber \\
    &
    \sum_{j=0}^{\min\{i,n'-k\}} \frac{i!(n'-k)!(-|T\alpha|)^{-2j}}{j!(i-j)!(n'-k-j)!}
    \sum_{j=0}^{\min\{i,n-k\}} \frac{i!(n-k)!(-|T\alpha|)^{-2j}}{j!(i-j)!(n-k-j)!}
    \ketbra{n'}{n}
\end{align}
\end{widetext}

In the limit $|R'| \to 1, T\alpha = \mathrm{const.}$ [discussed in Sec.~\ref{MZI section}],
only the summand corresponding to $k=0$ survives in [Eq.~\ref{Mi}] and $M_i$ reduces to a projector onto the state vector $\ket{i,T\alpha} = \hat D(T\alpha)\ket{i}$ of a generalised coherent state (GCS):
\begin{equation}
M_i \to \hat D^\dagger(T\alpha) \ketbra{i} \hat D(T\alpha)
\end{equation}
and we obtain a projective measurement as the limiting case.
(See Appendix~\ref{Derivation of POVM M$_i$} for details). 

\begin{figure*}[ht]
		\centering
		\includegraphics[width=\linewidth]{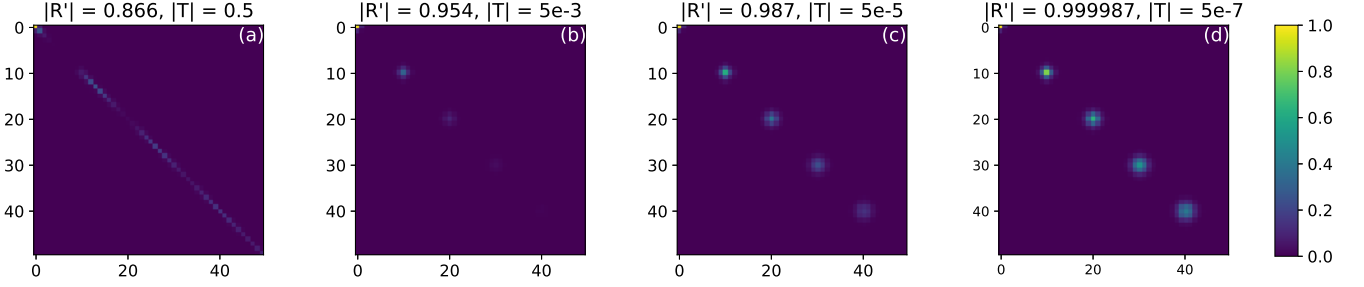}
		\caption{\textbf{POVM element $M_i$:} Numerically generated plot for the matrix representation (having dimensions $50\times50$) of the POVM $|M_i|$ for $i = 0$ to $50$ in gaps of $10$. The value of $\alpha = 0.1/T$ such that $T\alpha$ is kept constant throughout the simulation. The matrix $M_i$ is generated for different values of reflectivity ($|R'|$) and transmittivity ($|T|$) as shown in the subfigures: (a) $|R'| = 0.866$, $|T| = 0.5$; (b) $|R'| = 0.954$, $|T| = 5\times10^{-3}$; (c) $|R'| = 0.987$, $|T| = 5\times10^{-5}$ and (d) $|R'| = 0.999987$, $|T| = 5\times10^{-7}$.}
		\label{Mi_mat}
\end{figure*}

\begin{figure*}[ht]
		\centering
		\includegraphics[width=\textwidth]{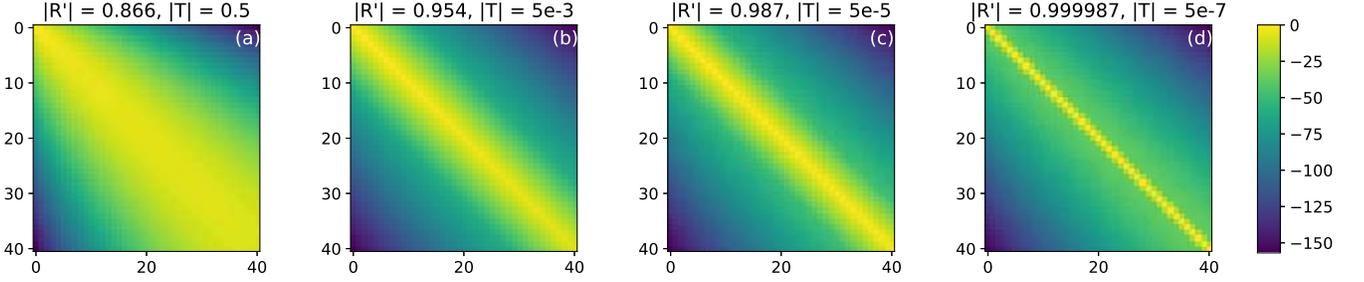}
		\caption{\textbf{Overlap between $M_i$ and $M_j$:} Numerically generated plot (\textbf{in logarithmic scale}) for the overlap between POVMs $M_i$ and $M_j$ with $i,j \in [0, 40]$. The value of $\alpha = 0.1/T$ such that $T\alpha$ is kept constant throughout the simulation. The matrix $M_i$ is generated for different values of reflectivity (R') and transmittivity (T) as shown in the subfigures: (a) $|R'| = 0.866$, $|T| = 0.5$; (b) $|R'| = 0.954$, $|T| = 5\times10^{-3}$; (c) $|R'| = 0.987$, $|T| = 5\times10^{-5}$ and (d) $|R'| = 0.999987$, $|T| = 5\times10^{-7}$.}
		\label{gram_mat}
\end{figure*}

In Fig.~\ref{Mi_mat} we have shown in one plot the numerically obtained matrix representations of $M_{i}$ ($i \in \{0,10,\dots,40\}$) for different values of $|R'|$ and $|T|$ of the MZI setup. From Figs.~\ref{Mi_mat} (a) to (d), the
values of $R'$ and $T$ slowly approach the limit of $|R'| = |R| \xrightarrow[]{} 1$, $T \xrightarrow[]{} 0$ and $\alpha \xrightarrow[]{} \infty$ under the condition that $T\alpha$ remains constant.


Next we have numerically estimated the overlap $\braket{M_j}{M_i}_{HS} = Tr(M_j^\dagger M_i)$ between the effects of $M_i$ and $M_j$ for each $i, j \in [0,40]$, estimating the effect's operators by  $52\times52$ matrices. We gather overlaps into a square matrix, being the (left-upper block of a) Gram matrix of the POVM. We have plotted the absolute values of its entries in the logarithmic scale 

[see Figs.~\ref{gram_mat} (a)-(d)].
We observe that for $|R'|=0.999987$ and $|T|=5\times10^{-7}$, we obtain an almost diagonal matrix, as expected for 
almost orthonomal operators $M_i$ approximating
the projective measurement.

\section{Maassen-Uffink Uncertainty Principle}\label{MU section}

In the previous section, we described the POVMs associated with the measurement made by a photon number detector at one arm of the MZI setting. While the MZI setup realizes the displacement operator $\hat D (\beta)$ under certain limits [discussed in Sec.~\ref{MZI section} and~\ref{POVMs}], the setup MZI + detector measures the observable $\hat D^\dagger (\beta) \hat N \hat D (\beta)$. We would like to comment now on the uncertainty relation between two such observables for two different values of $\beta$.

The Maassen-Uffink uncertainty principle~\cite{maassen1988generalized} deals with entropic uncertainties relying on Shannon entropy as a measure of uncertainty. The probability distributions for any quantum state $\ket{\psi}$ w.r.t two observables $A$ and $B$ having sets of eigenvectors ${\ket{a_j}}$ and ${\ket{b_j}}$ are $p=\abs{\bra{a_j}\ket{\psi}}^2$ and $q=\abs{\bra{b_j}\ket{\psi}}^2$, respectively. The Shannon entropy corresponding to any general probability distribution $x=(x_1,...,x_N)$ is given as $H(x)=-\sum_j x_j\log_2x_j$. For an $N$-dimensional Hilbert space, the Maassen-Uffink uncertainty principle is given as,

\begin{equation}\label{Maassen Uffink uncertainty}
    H(p) + H(q) \geq -2\log_2c
\end{equation}
where $c=\max_{j,k}\abs{\bra{a_j}\ket{b_k}}$. The right-hand side of [Eq.~\ref{Maassen Uffink uncertainty}] is independent of $\ket{\psi}$, \textit{i.e.}, the state of the system. Thus, non-trivial information is gathered about the probability distributions $p$ and $q$ from this relation, provided $c<1$.  

In the context of our problem, first we need to estimate the lower bound in [Eq.~\ref{Maassen Uffink uncertainty}]. 
The observables $\hat D^\dagger 
(\beta_1) \hat N \hat D (\beta_i)$, $i \in \{1,2\}$ has eigenbases $\{ D^\dagger(\beta_i) \ket{n} \}$ respectively. We want to find the maximum of $|\bra{m} D(\beta_1) D^\dagger(\beta_2) \ket{n}| = |\bra{m} D(\beta_1 - \beta_2) \ket{n}|$ over $n,m$. Let us provide the notation $\beta = \beta_1 - \beta_2$. 
The displacement operator $\hat D (\beta)$ acting on a state vector $\ket{n}$ produces a state known as a generalised coherent state (GCS)~\cite{boiteux1973semicoherent,philbin2014generalized,de1990properties}, which can be decomposed in the occupancy number basis:


$$\ket{n,\beta}=\hat{D}(\beta)\ket{n}=\sum_{k=0}^{\infty}C_{n,k}\ket{k}$$, 
where 
\begin{equation}
C_{n,k}=e^{-\abs{\beta}^2/2}\sum_{i=0}^{min(n,k)}\frac{\sqrt{n!}(-\beta^*)^{n-i}}{\sqrt{i!}(n-i)!}\frac{\sqrt{k!}(\beta)^{k-i}}{\sqrt{i!}(k-i)!}
\label{Cnk}
\end{equation}

[see Appendix~\ref{Generalized Coherent States}]. 



Now, numerically analysing $C_{n,k}$ 
[Eq.~\ref{Cnk}] for many values of $\beta$, 
we have obtained the following observation:
\begin{conjecture}
The maximum of $|C_{n,k}|$ is realised for $n=0$ (or $k=0$).
\end{conjecture}

\begin{figure}[ht]
		\centering
		\includegraphics[width=\linewidth]{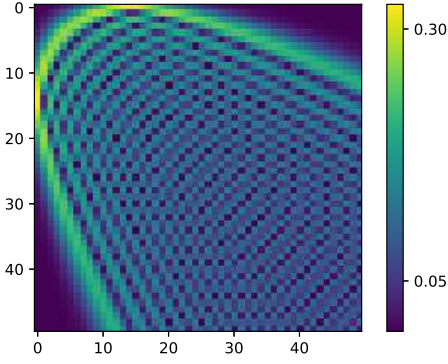}
		\caption{\textbf{Displacement Operator: Matrix Elements} Numerically generated plot for the absolute values of matrix elements ($\abs{C_{n,k}}=\abs{\bra{n}\hat{D}(\beta)\ket{k}}$) of the displacement operator ($\hat{D}(\beta)$) with $\beta=3.8$.}
		\label{Displacement Operator for gcs}
\end{figure}

[see Fig.~\ref{Displacement Operator for gcs}]. 

Using the above conjecture, we proceed analytically. It is straightforward to observe, that the sequence $C_{0,k}$ (coefficients of a 
coherent state in the occupancy number basis) satisfies the following recurrence relation: $C_{0,k}=\frac{\beta}{\sqrt{k}}C_{0,k-1}$ and we easily observe, that $\max_k |C_{0,k}|$ is at $k=\abs{\beta}^2$
(rounded to one of the nearest integers).

Hence we get,

\begin{equation}\label{Cnk max n,k}
    \max_{n,k} | C_{n,k} | = | C_{0,\abs{\beta}^2} | = e^{-\abs{\beta}^2/2}\frac{|\beta|^{\abs{\beta}^2}}{\sqrt{\Gamma(\abs{\beta}^2+1)}}
\end{equation}

Applying Stirling's formula to  $\Gamma(\abs{\beta}^2+1)$, \textit{i.e.}, $\Gamma(\abs{\beta}^2+1) > \sqrt{2\pi\abs{\beta}^2}\bigg(\frac{\abs{\beta}^2}{e}\bigg)^{\abs{\beta}^2}$ we have,

\begin{equation}\label{Gnk stirling}
     \abs{C_{0,\abs{\beta}^2}} <     e^{-\abs{\beta}^2/2}\frac{\abs{\beta}^{\abs{\beta}^2}}{\bigg(\sqrt{2\pi\abs{\beta}^2}\bigg(\frac{\abs{\beta}^2}{e}\bigg)^{\abs{\beta}^2}\bigg)^{1/2}}
\end{equation}

On simplifying the above equation, we arrive at,

\begin{equation}\label{simplified Gnk}
    \abs{C_{0,\abs{\beta}^2}} < \frac{1}{\sqrt[4]{2\pi\abs{\beta}^2}}.
\end{equation}


hence $c < \left(2\pi\abs{\beta}^2\right)^{-1/4}$ and the [Eq.~\ref{Maassen Uffink uncertainty}] gives us:

\begin{equation}\label{MU our problem}
    H(p) + H(q) \geq \frac 12 \log_2(2\pi\abs{\beta_1 - \beta_2}^2)
\end{equation}

where $p_i=\abs{\bra{i}\hat{D}(\beta_1)\ket{\psi}}^2$, $q_i=\abs{\bra{i}\hat{D}(\beta_2)\ket{\psi}}^2$ and $\abs{\beta}=\abs{\beta_1-\beta_2}$.
In a finite dimensional Hilbert space, the bound in the [Eq.~\ref{Maassen Uffink uncertainty}] is for a pair of observables having their eigenbases unbiased (related by a Hadamard unitary matrix) and cannot exceed $\log_2 d$, where $d$ is the dimension of the Hilbert space. In our case the dimension of the Hilbert space is infinite and the bound in [Eq. \ref{MU our problem}] is unbounded and grows with the moduli of the difference of the displacements.

\section{CHSH Inequality}\label{CHSH section}

The violation of the CHSH inequality is seen as the experimental confirmation of the entangled nature of the concerned states~\cite{clauser1969proposed}. Therefore, in this section we theorise the observables for experimentally establishing a test for the entanglement of the multiphoton state ($\ket{\psi}$), described in the previous sections.

Experimental realization of multiphoton entanglement detection would require number-resolved measurements on the outcoming photonic wavepacket from the beam splitter. Till date, the best possible resolution for photon detection is restricted to measuring temporally spaced single photons~\cite{hadfield2009single}, \textit{i.e.}, identifying the number of photons in a single pulse is not yet possible. Therefore, when 
a detector is placed at the output port of the MZI setting, either zero or non-zero number of photons will be reported by the detector. Let us prescribe outputs $-1$ and $1$ to these possibilities. The related observable will be:

\begin{align}\label{Observable A_beta}
    A(\beta) &= (-1)\ketbra{\beta}{\beta}+(+1)(\mathds{I}-\ketbra{\beta}{\beta})\nonumber\\
     &= \mathds{I}-2\ketbra{\beta}{\beta}
\end{align}

Here $\ket{-\beta}=\hat{D}^\dagger(\beta)\ket{0}$ is the vector corresponding to the measurement output of $-1$.


Let us assume that we have two such observables $A(\beta_1)$ and $A(\beta_2)$. For a state vector $\Psi$, the output statistics of both observables will be determined by two probabilities of getting an output of -1 for each of them:

\begin{equation}
p(A_i = -1 | \Psi) = |\braket{\Psi}{-\beta_i}|^2, \ i=1,2
\end{equation}


The output statistics is determined by the projection of $\Psi$ ($\Pi_V \Psi$) onto $V = \mathrm{span}\{\beta_1,\beta_2\}$. While $\Pi_V \Psi$ can have arbitrary norm $\le 1$, the effective Hilbert space must have at least one direction orthogonal to $V$, to project a normalised $\Psi$ onto $\Pi_V \Psi$ of desired norm. One orthogonal direction is enough to obtain it and hence the dimension of the effective Hilbert space for both observables is 3.

Let us fix an orthonormal basis of the effective Hilbert space ($\mathcal{H}$). Assuming that the displacement applied is $-\beta_1$ or $-\beta_2$ let, 
\begin{align}\label{basis vectors}
\ket{e_1} & = \ket{\beta_1}, \nonumber\\
\ket{e_2} & = \frac{\ket{\beta_2} - \braket{\beta_1}{\beta_2}\ket{\beta_1}}{\sqrt{1 - |\braket{\beta_1}{\beta_2}|^2}} \nonumber \\
&= \frac{\ket{\beta_2} - \braket{\beta_1}{\beta_2}\ket{\beta_1}}{\sqrt{1 - \exp(-|\beta_1-\beta_2|^2)}}
\end{align}
and let $\ket{e_3}$ be an arbitrary vector orthogonal to $\ket{\beta_1}$, $\ket{\beta_2}$.
Considering $\{\ket{e_1}, \ket{e_2}, \ket{e_3}\}$ as the basis for $\mathcal{H}$, the observables $A_1$, $A_2$ are represented by matrices:

\begin{equation}\label{Observable Matrix A0}
    A(\beta_1) = \begin{pmatrix}
       -1 & 0 & 0\\
        0 & 1 & 0\\
        0 & 0 & 1
        \end{pmatrix}
\end{equation}

\begin{equation}\label{Observable Matrix Aalpha}
    A(\beta_2) = \begin{pmatrix}
        1-2E & -2\sqrt{E(1-E)} & 0\\
        -2\sqrt{E(1-E)} & -1+2E & 0\\
        0 & 0 & 1
        \end{pmatrix},
\end{equation}
where $E = \exp(-|\beta_1-\beta_2|^2)$.

Let us assume that we have a source producing copies of a two-mode, multiphoton state. Consider an experiment, where these two modes become spatially separated and for each state from the pair, simultaneous measurements are performed in two distant laboratories. First laboratory chooses the displacement in the MZI setup to be $-\beta_1$ or $-\beta_2$ randomly, measuring the observables $A(\beta_1)$ and $A(\beta_2)$. Similarly, the second laboratory chooses randomly the displacement in the MZI setup to be $-\beta_3$ or $-\beta_4$, measuring observables $A(\beta_3)$ and $A(\beta_4)$. Both parties then perform a Bell-like experiment, similar to ~\cite{bell1964einstein, clauser1969proposed}.

Each party possesses a pair of dichotomic observables with outcomes $\pm 1$, hence the celebrated CHSH inequality:
\begin{eqnarray*}
| \mathbb{E} \big(
A(\beta_1)\otimes A(\beta_3) +A(\beta_2)\otimes A(\beta_3) \\ +A(\beta_1)\otimes A(\beta_4) -A(\beta_2)\otimes B(\beta_4) \big) | \le 2
\end{eqnarray*}
should hold for classically correlated states. The expression on the left-hand side is a non-local observable. Its expected value is reconstructed from local measurements. If the absolute value of its expected value exceeds 2, the state of two modes must be entangled.

The CHSH inequality can be violated if the maximal eigenvalue of the non-local observable it deals with, exceeds 2. The maximum eigenvalue of the observable is equal to:
\begin{equation}\label{upperbound}
    \lambda_{max} = 2\sqrt{1+4\sqrt[4]{E_1(1-E_1)}\sqrt[4]{E_2(1-E_2)}},
\end{equation}
where $E_1 = \exp(-|\beta_1 - \beta_2|^2)$, $E_2 = \exp(-|\beta_3 - \beta_4|^2)$. The above expression attains its maximal value for $E_1 = E_2 = 1/2$, what corresponds to:
\begin{equation}
    |\beta_1 - \beta_2|^2 = 
    |\beta_3 - \beta_4|^2 = \ln 2.
\end{equation}
For such settings $\lambda_{max} = 2\sqrt{2}$, which is exactly the Tsirelson's bound for the standard CHSH inequality~\cite{cirel1980quantum}.

The entangled state, for which the CHSH inequality is maximally violated is a projector onto the state vector:

\begin{equation}\label{pure state}
    \Psi = \frac{1}{2\sqrt{2-\sqrt{2}}}\begin{pmatrix}
                        -1 \\
                        1-\sqrt{2}\\
                        0\\
                        1-\sqrt{2}\\
                        1\\
                        0\\
                        0\\
                        0\\
                        0
                        \end{pmatrix}
\end{equation}

The state lives in the two-qubit subspace of $\mathbb{C}^3 \otimes \mathbb{C}^3$. By calculating its partial trace one can check, that this is a maximally entangled state of two qubits. This is what we expect from a state maximising the violation of CHSH inequality.

Let us express the above state vector in terms of the state vectors $\ket{\beta_i}$. Using formulas [Eq.~\ref{basis vectors}] one obtains:
\begin{widetext}
\begin{align}\label{pure state coherent states}
    \ket{\Psi} = \frac{1}{2\sqrt{2-\sqrt{2}}}\Big\{
    &\big[ (1 - e^{i\phi_1} - \sqrt{2}) \ket{\beta_1} + \sqrt{2} \ket{\beta_2} \big]
    \otimes
    \big[ (1 - e^{i\phi_2} - \sqrt{2}) \ket{\beta_3} + \sqrt{2} \ket{\beta_4} \big]
    - 2(2-\sqrt{2}) \ket{\beta_1} \otimes \ket{\beta_3}
    \Big\}
\end{align}
\end{widetext}
where we have used the following:
\begin{eqnarray}
    \braket{\beta_1}{\beta_2} = \sqrt{E_1} e^{-\beta_2\beta_1^*+\beta_2\beta_1^*}, \\
    \braket{\beta_3}{\beta_4} = \sqrt{E_2} e^{-\beta_4\beta_3^*+\beta_4\beta_3^*},
\end{eqnarray}
substituting the maximising values: $E_1 = E_2 = \frac 12$ and introducing notations: $i \phi_1=-\beta_2\beta_1^*+\beta_2^*\beta_1$ and similarly, $i \phi_2=-\beta_4\beta_3^*+\beta_4^*\beta_3$. 

The above formula takes a particularly simple form if $\phi_1=\phi_2=0$:
\begin{align}
    \ket{\Psi} = \frac{1}{\sqrt{2-\sqrt{2}}}\Big\{
    & \big[ \ket{\beta_1} - \ket{\beta_2} \big]
    \otimes
    \big[ \ket{\beta_3} - \ket{\beta_4} \big] \nonumber \\
    - & (2-\sqrt{2}) \ket{\beta_1} \otimes \ket{\beta_3}
    \Big\}
\end{align}
For this condition to hold, we must have $\{\beta_1\beta_2^*, \beta_3\beta_4^*\} \in \mathds{R}$, i.e., the relative phases of $\beta_1$, $\beta_2$ and $\beta_3$, $\beta_4$ are $0$.

\section{Conclusion}

We have devised a scheme for detecting entanglement in multiphotonic states using entanglement witnesses based on MZI setups. First, we have shown that while a quantum beam splitter fed with a strong coherent laser beam can effectively displace an input quantum state, the MZI setup comprising 50:50 beam splitters and a small relative phase shift can actually implement this. For a many-photon input state, a generalised coherent state (GCS) is observed at one of the output ports. 


Next, we have derived the uncertainty associated with the measurement observable (output intensity) when two different  displacements are produced by the MZI setup. This uncertainty increases as a function of the difference between the displacements. Finally, we have introduced entanglement witnesses that obey the CHSH inequality for testing entanglement in two-mode multiphotonic states. We also show the structure of the entangled state that causes maximal violation of the CHSH inequality. It was found that such a such a state can be prepared using coherent states (which are in fact, close to classical states). 

However, note that certain restrictions are imposed on the bound of the CHSH inequality by the detector inefficiency.
It has been shown that if the detector efficiency falls down to $\geq \approx 85.4 \%$, the bound in the CHSH inequality rises to the Tsirelson's bound~\cite{larsson1998bell}. 

At the end, keep in mind that the MZI setup realises the displacement operator in the approximate way - in fact, there is a trace amount of entanglement between output ports. As the second port is not measured, on the first port a POVM measurment performed. The bigger $|\alpha|$, the closer we get to a projective measurment.

\appendix

\begin{widetext}
\section{Derivation of POVM M$_i$}\label{Derivation of POVM M$_i$}

Considering the output state generated by the MZI setup with input $\ket{\psi}_0 \otimes \ket{\alpha}_1$, (analogous to [Eq.~\ref{BS: Output state}] for the beam splitter output state):

	\begin{equation}
	    \ket{\psi}_0 \otimes \ket{\alpha}_1 \xrightarrow[]{BS} \ket{\Psi}_{out}
	    = \hat{D}_4(R\alpha) \hat{D}_5(T\alpha)\sum_{n=0}^{\infty}c_n\frac{(T'\hat{a}_{2}^\dagger + R'\hat{a}_{5}^\dagger)^n}{\sqrt{n!}} \ket{0}_4\otimes\ket{0}_5
	\end{equation}

where $c_n = \braket{n}{\psi}$. The photon annihilation operators corresponding to the output ports of the MZI setting are $\hat{a}_4$ and $\hat{a}_5$ [see Fig.~\ref{MZI setup}]. The projector on the state of this composite system is:

	\begin{equation}\label{Projector_state}
	    \rho = \hat{D}_4(R\alpha) \hat{D}_5(T\alpha)\sum_{n=0}^{\infty}c_n\frac{(T'\hat{a}_{4}^\dagger + R'\hat{a}_{5}^\dagger)^n}{\sqrt{n!}} \ketbra{0}{0}_4\otimes\ketbra{0}{0}_5\sum_{n'=0}^{\infty}c^*_{n'}\frac{(T'^*\hat{a}_{4} + R'^*\hat{a}_{5})^{n'}}{\sqrt{n'!}}\hat{D}^\dagger_4(R\alpha) \hat{D}^\dagger_5(T\alpha)
	\end{equation}
	
We can easily check that,

\begin{equation}\label{binomial_expansion}
    \frac{(T'\hat{a}_{4}^\dagger + R'\hat{a}_{5}^\dagger)^n}{\sqrt{n!}} \ket{0}_4\otimes\ket{0}_5 = \sum_{k=0}^{n} \sqrt{{n \choose k}} T'^k R'^{n-k} \ket{k}_4\otimes\ket{n-k}_5
\end{equation}

Therefore, [Eq.~\ref{Projector_state}] reduces to:

\begin{align}
    \rho = \sum_{n,n'}c_nc^*_{n'}\sum_{k=0}^{n}\sum_{k'=0}^{n'}&\sqrt{{n \choose k}{n' \choose k'}} (T')^k (R')^{n-k}(T'^*)^{k'} (R'^*)^{n'-k'}\nonumber \\
    &\times\hat{D}_4(R\alpha) \hat{D}_5(T\alpha)\ketbra{k}{k'}_4\otimes\ketbra{n-k}{n'-k'}_5\hat{D}^\dagger_4(R\alpha) \hat{D}^\dagger_5(T\alpha)
\end{align}

The displaced photon number state is observed at the output port corresponding to the annihilation operator $\hat{a}_5$. Taking the partial trace over the first subsystem we have:

\begin{equation}
    \rho_5 = Tr_4(\rho) = \sum_{n,n'}c_nc^*_{n'}\sum_{k=0}^{min\{n,n'\}}\sqrt{{n \choose k}{n' \choose k}}\abs{T'}^{2k}(R')^{n-k}(R'^*)^{n'-k}\hat{D}_5(T\alpha)\ketbra{n-k}{n'-k}_5\hat{D}^\dagger_5(T\alpha)
\end{equation}

where we have used cyclic property of trace and $\hat{D}_4(R\alpha)^\dagger\hat{D}_4(R\alpha) = \mathds{I}$ Now, detecting $i$ photons from such a state can be represented by,

\begin{align}
    Tr(\ketbra{i}{i}\rho_5) =  Tr\Bigg(e^{-\abs{T\alpha}^2}\sum_{n,n'}\sum_{k=0}^{min\{n,n'\}}
    &
    \frac{\abs{T'}^{2k}(R')^{n-k}(R'^*)^{n'-k}\sqrt{n!n'!}}{k!\sqrt{(n-k)!(n'-k)!}}
    \nonumber\\&
    \sum_{j=0}^{\min\{i,n'-k\}}\frac{\sqrt{i!}(T^*\alpha^*)^{i-j}}{\sqrt{j!}(i-j)!}\frac{\sqrt{(n'-k)!}(-T\alpha)^{n'-k-j}}{\sqrt{j!}(n'-k-j)!}
    \nonumber\\&
    \times \sum_{j'=0}^{\min\{i,n-k\}}\frac{\sqrt{i!}(T\alpha)^{i-j'}}{\sqrt{j'!}(i-j')!}\frac{\sqrt{(n-k)!}(-T^*\alpha^*)^{n-k-j'}}{\sqrt{j'!}(n-k-j')!}\ketbra{n'}{n}\ketbra{\psi}{\psi}\Bigg)
\end{align}

using the cyclic property of trace and calculating  $\bra{i}\hat{D}_5(T\alpha)\ket{m}=\bra{-T\alpha}\ket{m}$ with $m \in \{(n-k),(n'-k)\}$ from [Eq.~\ref{gcs decomposed}]. We have also taken into account that $c_n = \braket{n}{\psi}$. 

Therefore, the POVM element corresponding to measuring $i$ photons at the detector end is:

\begin{align}\label{Mi appendix}
    M_i = e^{-\abs{T\alpha}^2} \frac{|T\alpha|^{2i}}{i!}
    \sum_{n,n'=0}^{\infty}
    &
    (-R'^*T\alpha)^{n'} (-R'^*T\alpha)^{*n}
    \sum_{k=0}^{\min\{n,n'\}} \frac{|R'\alpha|^{-2k}}{k!} \frac{\sqrt{n'!n!}}{(n'-k)!(n-k)!}
    \nonumber \\
    &
    \sum_{j=0}^{\min\{i,n'-k\}} \frac{i!(n'-k)!(-|T\alpha|)^{-2j}}{j!(i-j)!(n'-k-j)!}
    \sum_{j=0}^{\min\{i,n-k\}} \frac{i!(n-k)!(-|T\alpha|)^{-2j}}{j!(i-j)!(n-k-j)!}
    \ketbra{n'}{n}
\end{align}

where we have used the Stokes' law $R'T^*+R^*T' = 0$. Moreover, $M_i$ is takes into account moduli of R', T and T'. So their relative phases can be neglected. In the limits $T \xrightarrow{} 0$, $|R'| = |R| \xrightarrow{} 1$ and $\alpha \xrightarrow{} \infty$ but $T\alpha$ remains constant, only $k = 0$ term dominates in the summation. So the form of $M_i$ [from Eq.~\ref{Mi appendix}] in such a case is:

\begin{equation}
M_i = \Bigg(\frac{\exp\left(-\frac{|T\alpha|^2}{2}\right)}{\sqrt{i!}}
\sum_{n'=0}^{\infty}\sum_{j=0}^{\min\{i,n'\}}
\frac{i!(T\alpha)^{n'-j}(-T^*\alpha^*)^{i-j}}
{j!(i-j)!(n'-j)!}\sqrt{n'!}\ket{n'}\Bigg)\times h.c.
\end{equation}

On comparing the above with [Eq.~\ref{gcs decomposed}] (upto relabeling of indices and changing the summation variable), we see that $M_i$ reduces to a projector onto the generalised coherent state $\ketbra{i,-T\alpha}{i,-T\alpha}$. 

\section{Generalized Coherent States}\label{Generalized Coherent States}

The displacement operator acting on an $n$-photon state gives rise to generalized coherent states (CGS) given as,

\begin{equation}\label{gen coherent states}
\ket{n,\beta} = \hat{D}(\beta)\ket{n}
\end{equation}

Now, applying the Baker-Campbell-Hausdorff formula to the displacement operator,
we can expand the above expression as
follows, to obtain the exact functional form of $\ket{n,\beta}$,

\begin{equation}\label{gcs with disp expanded}
    \ket{n,\beta} = e^{-\abs{\beta}^2/2}e^{\beta \hat a^\dagger}e^{-\beta^* \hat a}\ket{n}
\end{equation}

Using the Taylor expansion of 
exponents
we get:

\begin{equation}\label{gcs taylor expanded}
    \ket{n,\beta} = e^{-\abs{\beta}^2/2}
    \sum_{j=0}^{\infty}\sum_{i=0}^{\infty}
    \frac{(\beta \hat a^\dagger)^j}{j!}
    \frac{(-\beta^*\hat a)^i}{i!}
    \ket{n}
\end{equation}
\\
The powers of creation/annihilation operators act on occupancy number states as follows:

\begin{equation}\label{creation and annihilation operator props}
\begin{gathered}[b]
    \hat a^{_\dagger l}\ket{m} = \sqrt{\frac{(m+l)!}{m!}}\ket{m+l} \\
    \hat a^{l}\ket{m} = \sqrt{\frac{m!}{(m-l)!}}\ket{m-l}
\end{gathered}
\end{equation}

In [Eq.~\ref{gcs taylor expanded}], we obtain

\begin{align}\label{gcs expression factorials}
\ket{n,\beta} = & e^{-\abs{\beta}^2/2}
\sum_{j=0}^{\infty}
\sum_{i=0}^{n}
\frac{(\beta)^j}{j!}
\frac{(-\beta^*)^i}{i!}
\sqrt{\frac{(n-i+j)!}{(n-i)!}}
\sqrt{\frac{n!}{(n-i)!}}
\ket{n-i+j}
\nonumber\\ = & 
\frac{e^{-\abs{\beta}^2/2}}{\sqrt{n!}}
\sum_{k=0}^{\infty}
\sum_{i=\max\{0,n-k\}}^{n}
\frac{(-\beta^*)^i}{i!}
\frac{(\beta)^{k-n+i}}{(k-n+i)!}
\frac{n!}{(n-i)!}
\sqrt{k!}\ket{k}
\nonumber\\ =& 
\sum_{k} C_{n,k}\ket{k}
\end{align}
where the reparametrisation has been done introducing a new variable $k=n-i+j$

and 
the summation limits has been changed accordingly, as the Fig.~\ref{GCS_reparam} explains.

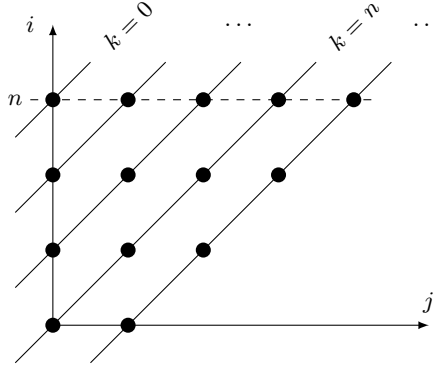
\begin{figure}[H]
	\centering
    \begin{tikzpicture}
        \draw[->,>=latex] (0,0) -- (0,4);
        \draw[->,>=latex] (0,0) -- (5,0);
        \fill (0,0) circle (.1);
        \fill (0,1) circle (.1);
        \fill (0,2) circle (.1);
        \fill (0,3) circle (.1);
        \fill (1,0) circle (.1);
        \fill (1,1) circle (.1);
        \fill (1,2) circle (.1);
        \fill (1,3) circle (.1);
        \fill (2,1) circle (.1);
        \fill (2,2) circle (.1);
        \fill (2,3) circle (.1);
        \fill (3,2) circle (.1);
        \fill (3,3) circle (.1);
        \fill (4,3) circle (.1);
        \draw (-.5,2.5) -- (.5,3.5);
        \draw (-.5,1.5) -- (1.5,3.5);
        \draw (-.5,.5) -- (2.5,3.5);
        \draw (-.5,-.5) -- (3.5,3.5);
        \draw (.5,-.5) -- (4.5,3.5);
        \node at (-.3,4) {$i$};
        \node at (5,.3) {$j$};
        \node[rotate=45] at (1,4) {$k=0$};
        \node at (2.5,4) {$\dots$};
        \node[rotate=45] at (4,4) {$k=n$};
        \node at (5,4) {$\dots$};
        \draw[dashed] (-.3,3) -- (4.3,3);
        \node at (-.5,3) {$n$};
    \end{tikzpicture}
    \caption{\textbf{Reparametrisation of the summation area in (\ref{gcs expression factorials})} The new variable $k$ takes nonnegative values. For a given $k$, the variable $i$ takes values from the range $\{0, \dots, n\}$, except for $k < n$, when the range of $i$ is $\{n-k, \dots, n\}$.
	}
	\label{GCS_reparam}
\end{figure}

One can easily check that the above expression can be reduced to a form involving associated Laguerre polynomials, as introduced in earlier papers~\cite{cahill1969ordered, philbin2014generalized}.
However, if one needs to generate the whole matrix of displacement operator, a slightly different representation of $C_{n,k}$ will be more convenient. 
After a reparametrisation by $i \mapsto n-i$, one can express
[Eq.~\ref{gcs expression factorials}] as follows

\begin{align}\label{gcs decomposed}
C_{n,k} &= 
e^{-\abs{\beta}^2/2}
\sum_{i=0}^{\min\{n,k\}}
\frac{(-\beta^*)^{n-i}}{(n-i)!}
\frac{(\beta)^{k-i}}{(k-i)!}
\frac{\sqrt{n!k!}}{i!}
\nonumber\\ &
= e^{-\abs{\beta}^2/2}
\sum_{i=0}^{\min\{n,k\}}
\frac{\sqrt{n!}(-\beta^*)^{n-i}}{\sqrt{i!}(n-i)!}
\frac{\sqrt{k!}(\beta)^{k-i}}{\sqrt{i!}(k-i)!}
\nonumber\\ &
= e^{-\abs{\beta}^2/2}
\braket{u_n(\beta^*)}{u_n(-\beta^*)},
\end{align}
where $\ket{u_n(\beta)} = \sum_{i=0}^{n} \frac{\sqrt{n!}(\beta)^{n-i}}{\sqrt{i!}(n-i)!} \ket{i}$. Hence the matrix of the displacement operator in the occupancy eigenbasis can be decomposed as
\begin{align}\label{gcs with dot product}
D(\beta) &= e^{-\abs{\beta}^2/2} \vb U(\beta^*)^\dagger \vb U(-\beta^*),
\end{align}
where columns of $\vb U(\beta)$ are the subsequent vectors $u_n(\beta)$. One can check, that $\vb U(-\beta^*)$ is a matrix representation of $\exp(-\beta^* \hat a)$. Hence [Eq.~\ref{gcs with dot product}] is a matrix representation of operator equation [Eq.~\ref{gcs with disp expanded}].

\end{widetext}

%


\end{document}